# Dark Energy and Dark Matter in a Superfluid Universe[1]

Kerson Huang

Massachusetts Institute of Technology, Cambridge, MA , USA 02139 and
 Institute of Advanced Studies, Nanyang Technological University, Singapore 639673

## Abstract
The vacuum is filled with complex scalar fields, such as the Higgs field. These fields serve as order parameters for superfluidity (quantum phase coherence over macroscopic distances), making the entire universe a superfluid. We review a mathematical model consisting of two aspects: (a) emergence of the superfluid during the big bang; (b) observable manifestations of superfluidity in the present universe. The creation aspect requires a self-interacting scalar field that is asymptotically free, i.e., the interaction must grow from zero during the big bang, and this singles out the Halpern-Huang potential, which has exponential behavior for large fields. It leads to an equivalent cosmological constant that decays like a power law, and this gives dark energy without "fine-tuning". Quantum turbulence (chaotic vorticity) in the early universe was able to create all the matter in the universe, fulfilling the inflation scenario.  In the present universe, the superfluid can be phenomenologically described by a nonlinear Klein-Gordon equation. It predicts halos around galaxies with higher superfluid density, which is perceived as dark matter through gravitational lensing. In short, dark energy is the energy density of the cosmic superfluid, and dark matter arises from local fluctuations of the superfluid density

## 1. Overview

Physics in the twentieth century was dominated by the theory of general relativity on the one hand, and quantum theory on the other. General relativity gives us insight into the expanding universe and the big bang, while quantum theory gives u superfluity, (which expresses a coherence of the quantum phase over macroscopic distances), and the dynamical vacuum with fluctuating quantum fields. These facets will come together to offer explanations to the puzzles of our time: dark energy and dark matter.

The idea is that the vacuum is filled with a complex scalar field, such as the Higgs field, making the entire universe a superfluid. All astrophysical processes take place in this superfluid. Dark energy can be identified as the energy density of the superfluid, and dark matter is the manifestation of fluctuations of the superfluid density.

We first review relevant backgrounds in cosmology, superfluidity, and quantum field theory, particularly renormalization, and then describe the synthesis into a mathematical cosmological theory [1-3]. For general background on cosmology and general relativity, we refer to [4,5], and for background on quantum field theory and particle physics we refer to [6,7].

## 2. Background review

### a. Dark energy and dark matter

Hubble's law states that galaxies move away from us with speeds proportional to their distances from us. Since our location is not special, this implies that all galaxies move away from each other in this manner. Through general relativity, we understand that this means that the fabric of the universe is expanding, that such galactic motion may be likened to points on a balloon being blown up. The fractional rate of increase of the radius of the universe a(t), called the Hubble parameter, has the present value

$$H \equiv \frac{\dot{a}}{a} \approx \frac{1}{1.5 \times 10^{10} \text{ yr}}$$
(1)

Extrapolating the expansion backwards in time, one expects to arrive at a singular point in time when the universe was created, whimsically referred to as the "big bang". Fig.1 shows a plot of galactic distance vs. galactic red shift (a measure of velocity) [8]. On this plot, Hubble's law corresponds to the straight line. We can see that the nearby galaxies follow this law pretty well, but for larger distances the data deviate from the linear law. Galaxies at an earlier epoch were moving more slowly than expected, and this indicates that the expansion of the universe proceeds at an accelerated rate. One attributes the acceleration to an unseen energy in the universe called the "dark energy".

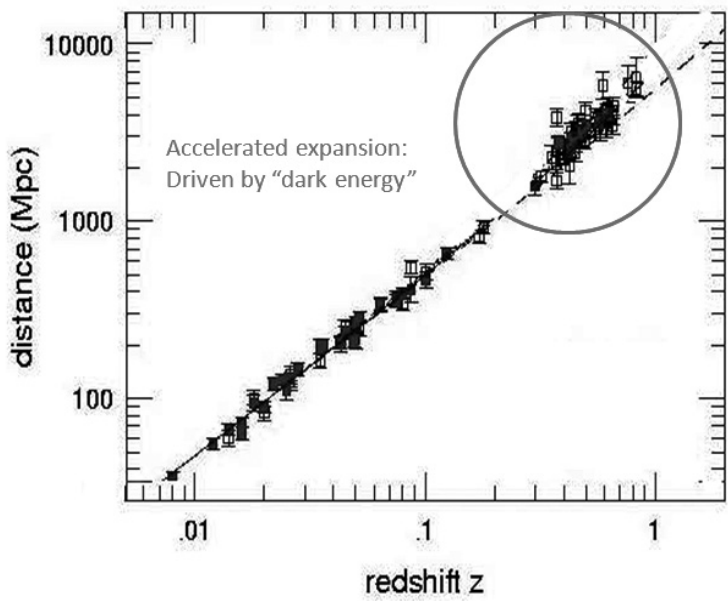

Fig1. Deviations from Hubble's law (straight line) indicate an accelerated expansion of the universe, leading to the postulate of "dark energy" as driving agent.

The idea of dark matter was originally suggested by the observation that the velocity of dust around a rotating galaxy does not drop off with distance, as one expects on the basis of the visible mass of galaxy [9], as illustrated in Fig.2. This leads to the hypothesis that there is an unobservable "dark matter" component of the galaxy that only has gravitational interactions. More recently, gravitational lensing reveals halos around galaxies that have gravitational but no electromagnetic interactions [10], as shown in Fig.3. This is the most convincing evidence for dark matter to-date.

It is estimated [11] that dark energy makes up 70% of the universe's total energy, and dark matter, 25%. Particle physicists speculate that dark matter consists of unknown elementary particles called WIMPs (weakly interacting massive particles), but they have eluded particle detectors so far [12].

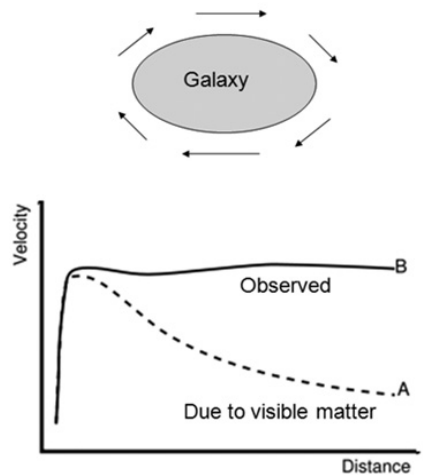

Fig2. Dust around a rotating galaxy has higher velocity then expected from that caused by the visible mass, prompting speculations of "dark matter" components.

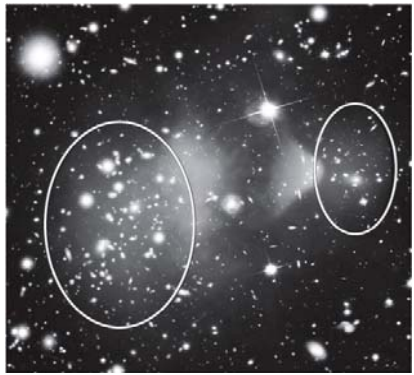

Fig3. Collision of two galaxy clusters (the bullet cluster) show trailing halos (in circles) detectable only through gravitational lensing. These are to-date the most convincing evidence of dark matter.

### b. Superfluidity

Superfluidity is experimentally manifested by the apparent lack of viscosity of liquid helium below a critical temperature. From a theoretical point of view, it expresses a coherence of the quantum phase over macroscopic distances. In this sense, the superconductivity of metals is a form of superfluidity. Ginsburg and Landau, in their 1950 phenomenological theory [13], describe this coherence in terms of an order parameter represented by a complex scalar field:

$$\phi(x) = F(x)e^{i\sigma(x)} \qquad (2)$$

The superfluid velocity (the supercurrent in the case of superconductivity) is identified with the gradient of the phase $\nabla\sigma$. An important application is the Josephson effect, in which a phase difference between two adjacent superconductors leads to a supercurrent.

A quantized vortex in a superfluid is a flow around a line, called the vortex line, such that the phase σ changes by an integer multiple of 2π around any closed curve enclosing the vortex line:

$$\oint d\mathbf{s} \cdot \nabla\sigma = 2\pi n \quad (n = 0, \pm 1, \pm 2, \dots) \qquad (3)$$

The quantization of vorticity makes quantum vortices simpler than their counterparts in a classical fluid, because they are identical, like hydrogen atoms. A chaotic tangle of vortex lines produces quantum turbulence [14].

### c. The dynamical vacuum

According to quantum field theory, there are fluctuating quantum fields in the vacuum state. Those of the electromagnetic field lead to the Lamb shift in hydrogen, which splits the 2S-2P level degeneracy by $10^{-6}$ eV, and the anomalous magnetic of the electron, which makes its g factor deviate from the value 2 given by the Dirac equation: g-2 = $\alpha/2\pi$, where $\alpha \approx 1/137$ is the fine-structure constant.

The standard model of particle physics [7] introduces the Higgs field, which has complex components that fluctuate about a nonzero vacuum field φ(x). Its existence has found experimental support in the discovery of the field quantum, the Higgs boson [15]. This field endows mass mainly in the weak sector, in particular the masses of the intermediate vector bosons, through the spontaneous breaking of local gauge invariance (the so-called Higgs mechanism). This mechanism was first used by Ginsburg and Landau [13] to explain the Meissner effect in superconductivity, wherein the photon acquires mass in a superconductor equal to the inverse of the penetration depth.

We now know from the microscopic BCS [16] theory that the scalar field of Ginsburg and Landau corresponds to the wave function of Cooper pairs. But the phenomenological order parameter remains a useful tool, and is widely used in the description of superfluidity in liquid helium and cold atomic gases [17]. We must conclude that the Higgs field, and other complex vacuum fields needed in grand-unified theories, make the universe a superfluid, regardless of the ultimate microscopic origin of these fields.

### d. Renormalization

In quantum field theory, there are virtual processes whose momentum spectrum extends all the way to infinity. This spectrum must be cut off at some finite momentum Λ, for otherwise the high-momentum processes will lead to divergences in the theory. Besides, any theory we write down is unlikely to be valid at very high momenta, or small distances. How should one deal with this cutoff? Dyson [18] shows in QED (quantum electrodynamics) that radiative corrections "renormalize" the bare charge $e_0$ of the electron into the physical charge $e_1$ given by

$$e_1 = Z(\Lambda) e_0 \tag{4}$$

where $Z(\Lambda)$ is a renormalization constant that diverges when $\Lambda \to \infty$. In practice, one takes $e_1$ from experiments, thus "burying" the cutoff. This trick leads to the triumph of QED --- the successful calculation of the Lamb shift. The physical meaning of renormalization emerges only gradually, through the works of Gell-Mann and Low, Callen and Symanzik, Bogoliubov, and finally the synthesis by Wilson [19, 20].

In Wilson's theory, a high-momentum cutoff $\Lambda_0$ should be part of the definition of the system, and for an isolated system it is the only scale. To describe the system at a lower scale $\Lambda$, one "hides" the degrees of freedom between $\Lambda$ and $\Lambda_0$, as indicted in Fig.4. This is achieved through renormalization, whereby all coupling constants undergo transformations, in order to preserve the basic identity of the theory. These transformations form the RG (renormalization group), which is a representation of the group of scale transformations. This gives rise to an effective Lagrangian that depends on the scale. The appearance of the system changes with the scale, but not its basic identity. An analogy is viewing a painting at increasing magnifications. While its appearance changes from art, to grains of paint on canvass, to arrangements of atoms, the object being viewed is the same.

e. Scalar field

Consider a scalar field ϕ(x) with effective Lagrangian density

$$\mathcal{L} = \tfrac{1}{2}(\partial \phi)^2 - V(\phi)$$
$$V(\phi) = g_2 \phi^2 + g_4 \phi^4 + g_6 \phi^6 + \cdots \tag{5}$$

The coupling constants $g_n(\Lambda)$ are functions of the effective cutoff $\Lambda$, which defines the scale. The potential $V(\phi)$ is therefore a function of the cutoff. As $\Lambda$ changes, the effective Lagrangian traces out a trajectory -- the RG trajectory -- in the space of all possible Lagrangians, as illustrated in Fig.5. The point corresponding to $V(\phi) \equiv 0$ represents the massless free field. It is a fixed point of RG called the Gaussian fixed point, at which the system is scale-invariant, and $\Lambda = \infty$.

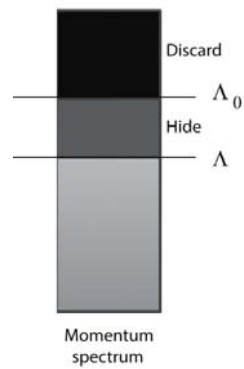

Fig4. The momentum spectrum of a quantum field theory must be cut off at some high momentum $\Lambda_0$. When we work at a lower momentum scale $\Lambda$, we "hide" the degrees of freedom in between by adjusting the coupling constants of the theory, through the process of renormalization.

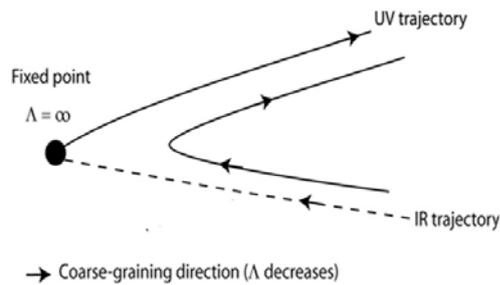

Fig5. RG (renormalization group) trajectories in the space of all possible Lagrangians. The effective Lagrangian traces out these trajectories under scale transformations, when the coupling constants undergo renormalizations. A fixed point is where the system is scale-invariant, and $\Lambda = \infty$. A UV (ultraviolet) trajectory goes away from the fixed point upon coarse-graining (decrease of $\Lambda$). An IR (infrared) trajectory, which goes into the fixed point, is a critical line along which $\Lambda = \infty$.

RG trajectories emanate from the Gaussian fixed point. Those trajectories that move away from the fixed point upon coarse-graining (decrease of $\Lambda$) are called UV (ultraviolet) trajectories, along which the system exhibits "asymptotic freedom". That is, the system approach a free theory in the reverse direction, when $\Lambda \rightarrow \infty$. The IR (infrared) trajectories go towards the fixed point under coarse-graining. Since $\Lambda$ decreases under coarse-graining, and it is infinite at the fixed point, we have $\Lambda = \infty$ on the entire trajectory. This makes it a critical line, along which the system is the same as that at the fixed point -- a free theory. A theory with finite $\Lambda$ cannot lie on the critical line. One can place it on an adjacent trajectory. When one moves it closer to the critical line, the system will approach a free field -- a phenomenon called "triviality". To avoid this, one either keeps Λ finite, or goes on a UV trajectory.

We should choose the original cutoff $\Lambda_o$ such that $\Lambda<\Lambda_o$. On a UV trajectory, this means $\Lambda_o=\infty$, corresponding to the fixed point. On an IR trajectory, $\Lambda_o=\infty$, for a massless free theory does not need a cutoff.

What is new in Wilson's theory is that the RG trajectory can go anywhere in the parameter space. It can break out into new dimensions not contained in the original Lagrangian. In contrast, the old formulation only admits "renormalizable" theories that are self-similar under scale transformations. That is, they remain in the subspace defined by the bare Langrangian. Wilson removes this restriction.

### f. Halpern-Huang potential

RG analyses show that the only asymptotically free scalar potential, near the Gaussian fixed point, is the Halpern-Huang potential [21-23], which for a scalar field with N components φ={φ$_1$,…,φ$_N$} is given by

$$V(\phi) = c\Lambda^{4-b}\left[ M\left(-2+\frac{b}{2}, \frac{N}{2}, \frac{8\pi^2\phi^2}{\Lambda}\right) - 1 \right]$$

$$\phi^2 \equiv \sum_n \varphi_n^2$$

(6)

where c is an arbitrary complex constant, *b* is a real constant, with b>0 corresponding to asymptotic freedom, and b<2 yielding spontaneous symmetry breaking, (i.e., a minimum of the potential occurring at φ>0). The function M is a Kummer function, a hypergeometric function defined by the power series

$$M(p,q,z) = 1 + \frac{p}{q}z + \frac{p(p+1)}{q(q+1)}\frac{z^2}{2!} + \frac{p(p+1)(p+2)}{q(q+1)(q+2)}\frac{z^3}{3!} + \cdots$$

(7)

It behaves like an exponential at large fields:

$$M(p,q,z) \approx \Gamma(q)\Gamma^{-1}(p)z^{p-q}\exp z$$

(8)

The potential is valid to lowest order in $\Lambda^{-1}$. The most general asymptotically free potential is a linear combination of Halpern-Huang potentials. All polynomial potentials, including the popular φ⁴ theory, are "trivial". An example of the Halpern-Huang for b=1 is shown in Fig.6.

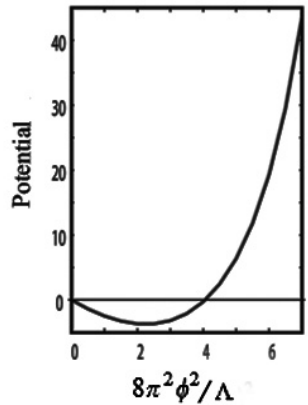

Fig6. Halpern-Huang scalar potential for b=1. It exhibits spontaneous symmetry breaking, is asymptotically free, and rises exponentially at large fields.

## 3. The big bang

Assuming that the vacuum scalar field emerges with the big bang, we must use the Halpern-Huang potential (or a linear combination thereof), for it should grow from zero as the length scale expands, i.e., it should be asymptotically free. Fig.7 illustrates the "Creation" in RG space. Just before the big bang, the scalar field was a massless free field at the Gaussian fixed point. At the big bang, it was infinitesimally displaced along some direction, onto some RG trajectory. If that happens to be an IR trajectory, then the system actually never left the fixed point. If that happens to be a UV trajectory, then V(φ) is a Halpern-Huang potential. As Λ decreases with the expanding universe, a possible universe evolves.

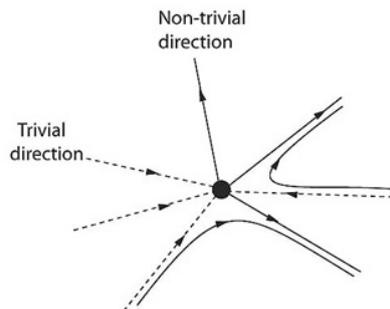

Fig7. At the big bang, the scalar field emerges from the Gaussian fixed point (massless free field), and moves along an RG trajectory as the universe expands. On a UV trajectory (non-trivial direction), it spawns a possible universe. On an IR trajectory (trivial direction), it goes back to the fixed point.

We can describe the big bang as a mathematical initial-value problem. We illustrate this with a real scalar field φ. The equations of motion are Einstein's equation (without cosmological term) plus the field equation of the scalar field:

$$\dot{H} = \frac{k}{a^2} - \dot{\phi}^2 + \frac{1}{3}a\frac{\partial V}{\partial a}$$
$$\ddot{\phi} = -3H\dot{\phi} - \frac{\partial V}{\partial \phi}$$

(9)

where G is the gravitational constant, and T μν is the energy-momentum tensor of the scalar field. Assuming that the universe is spatially uniform, we use the Robertson-Walker metric, which contains only one parameter a(t), the radius of the universe. The only scale for the scalar field is the high-momentum cutoff Λ, and the Halpern-Huang potential is a large-Λ expansion to lowest order in $\Lambda^{-1}$. Since there should be only one scale in the early universe, we tie it with a(t):

$$\Lambda = \frac{\hbar}{a}$$

(10)

This sets up a dynamical feedback loop: gravity cuts off the scalar field, which generates gravity. The independent variable are a(t), φ(t). With the Hubble parameter H≡$a^{-1}$da/dt, and in units ℏ=c=4πG=1, the equations reduce to

$$\dot{H} = \frac{k}{a^2} - \dot{\phi}^2 + \frac{1}{3}a\frac{\partial V}{\partial a}$$
$$\ddot{\phi} = -3H\dot{\phi} - \frac{\partial V}{\partial \phi}$$

(11)

plus a constraint equation

$$H = \left(\frac{2}{3}V + \frac{1}{3}\dot{\phi}^2 - \frac{k}{a^2}\right)^{1/2}$$

(12)

where k=0,±1 is the curvature parameter. The term (a/3)∂V/∂a in the first equation is absent in conventional theories where the potential V is independent of cutoff. Here, it arises from the trace anomaly [24] of Tμν, a renormalization effect in quantum field theory, and is needed for the preservation of the constraint. As a consequence, the constraint has to be enforced only once, in the beginning, where it restricts initial data to a small subset. This restriction is crucial for our results. As we can see from (12), a=0 is ruled out as an initial condition. We therefore start from a time after the big bang, but still within the Planck time of $10^{-43}$s. We assume that at this time the vacuum scalar field is already in place, i.e., φ(0)≠0. Physically speaking, the universe might have started at such a high temperature that the scalar system was in an uncondensed (symmetry-restored) phase, and that it rapidly cooled down to our initial state; but such scenarios are beyond the scope of our study.

We can obtain numerical solutions for a range of parameters in V(φ), and a range of initial conditions. A typical solution is shown in Fig.8. The salient feature is that the Hubble parameter decreases like a power $t^{-p}$ for large times, and this implies that the radius of the universe expands like exp $t^{-p}$ :

$$H \sim t^{-p}$$
$$a \sim \exp t^{1-p}$$
(13)

where p depends on parameters in the potential and on initial conditions. This gives dark energy, since the universe expands at an accelerated rate. Although we did not explicitly include a cosmological constant in Einstein's equation, the initial value of H gives an effective cosmological constant $H^2$, which decreases with time like $t^{2p}$. The initial value could be of order unity in Planck units, and the power-law decay would reduce it to the very small present value in the course of 15 billion years, and this solves the so-called "fine-tuning problem". We may say that dark energy comes from V(φ), the energy density of the scalar field.

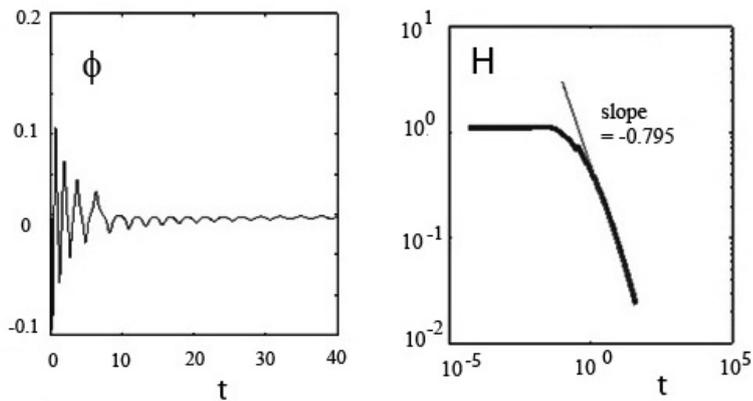

Fig8. Numerical solution of scalar cosmology with Halpern-Huang potential near the big bang. The parameters are k=-1, b=1, c=0.2. Other choices give qualitatively similar results. The field oscillates, and the Hubble parameter asymptotically decays like a power law. This is equivalently to a decaying cosmological constant, which gives dark energy without the 'fine tuning" problem.

Our model is not valid beyond the Planck era, because the Halpern-Huang potential is good only to lowest order in a=Λ$^{-1}$. Besides, density fluctuations and temperature effects would become important. Nevertheless, it is interesting to compare the power-law predictions with galactic red-shift data [8], in a phenomenological sense. In Fig.9, curve A is the theoretical prediction, with a definite exponent p. When we adjust p and/or other parameters, the curve will move vertically without change in shape, such as curve B. We can see that curve A well fits that low z data, but there seems to be an earlier epoch corresponding to curve B. We estimate that the crossover transition from B to A was completed about 7 billion years ago.

The power-law behavior of H has also been compared with WMAP data under the name "intermediate inflation" [25].

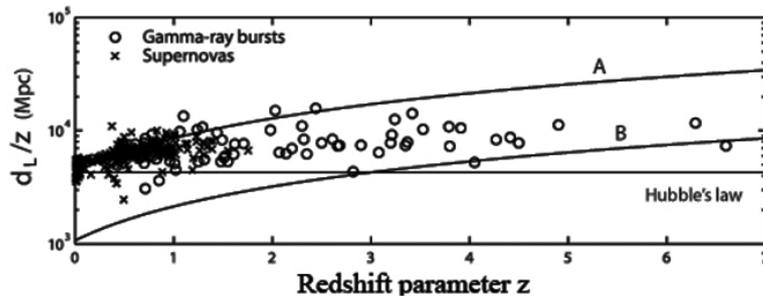

Fig9. Comparison of predictions of a power-law Hubble parameter with data on galactic red shift. Here, $d_L$ is the luminosity distance. The theoretical curve A can only be shifted up and down upon change of the power and/or model parameters. There seems to be a crossover transition from B to A that was completed about 7 billion years ago.

## 4. Matter creation

Conventional theories speak of an era of inflation [4] after the big bang, in which all matter was created during a rapid expansion of the universe. The universe then was small enough so the matter can "see" each other, and retain memory of a uniform density to the present day, when they have long fallen out of each other's horizons. These theories use classical scalar fields, which are quite different from our quantum field with a time-dependent cutoff. They invoke such things as "slow roll" and "reheating", which do not happen in our model. As seen in Fig.8, our scalar field undergoes rapid oscillations, rather then slow roll. The conventional theories have difficulty creating enough matter, and our model has the same problem. With a completely uniform scalar field, even with many components, our model has difficulty creating matter at a sufficiently high rate to satisfy the inflation scenario.

Another problem related to matter creation has to with the fact that matter brings in a new scale characterized by the nucleon mass of 1 GeV, which is to be compared with the Planck scale of $10^{18}$ GeV built into Einstein's equation. (This new scale arises spontaneously in QCD (quantum chromodynamics), through what is called "dimensional transmutation" [7]). The problem is how to mathematically decouple this nuclear scale from the Planck scale in the equations of motion.

These problems call for a new mechanism based on new physics, and we find it in the superfluidity of a complex scalar field, particularly quantum turbulence.

## 5. Quantum turbulence

The simplest vortex configuration is the vortex ring shown in Fig.10. It has a translation velocity normal to the ring, essentially inversely proportional to the radius. In the arbitrary vortex line shown in Fig.10, different points on the line moves at different velocities normal to, and inversely proportional to, the radius of curvature at that point. Thus, a vortex line moves incessantly with a self-driven writhing motion. Two vortex lines can cross and reconnect, as illustrated in Fig.11. After reconnection, there will appear two cusps on the final lines with theoretically zero radii of curvature, and they will spring away from each other will theoretically infinite speed. This will create two jets of energy in the superfluid, which could materialize as two jets of matter, if coupling exists between the superfluid and matter. Vortex reconnection is an efficient way to convert a large amount of potential energy into kinetic energy in a short time. A similar process, reconnection of magnetic flux lines, is responsible for solar flares, also illustrated in Fig.11.

Repeated reconnection of vortex lines can create a vortex tangle with fractal dimension, and this is quantum turbulence, which has been observed in superfluid helium [14, 26]. A computer simulation of the formation of quantum turbulence [27] is shown in Fig.12. In the vortex tangle, potential energy is converted to kinetic energy through vortex reconnections at a steady rate, we invoke this as the physical mechanism of matter creation in the early universe.

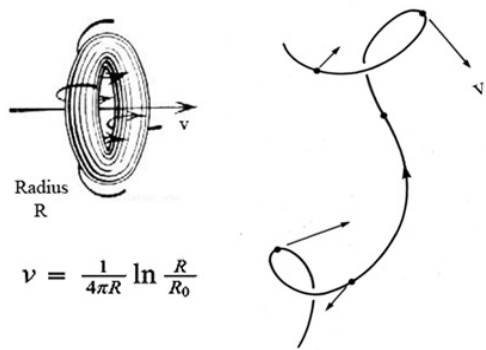

$$v = \frac{1}{4\pi R} \ln \frac{R}{R_0}$$

Fig10. A vortex line has a local velocity inversely proportional to the radius of curvature, and normal to it. It generally executes a complicated self-induced writhing movement.

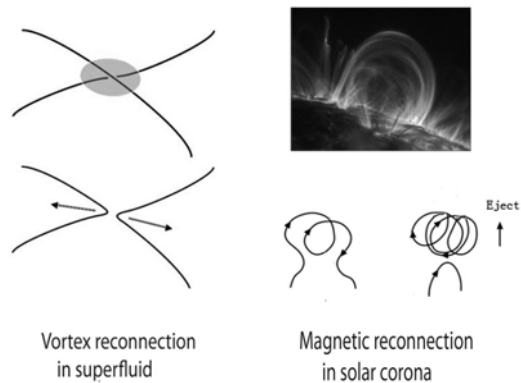

Fig11. Reconnection of two vortex lines produces on the final lines two cusps springing away from each other at theoretically infinite speed (because they have zero radius of curvature). This would create two jets of energy. The reconnection of magnetic flux lines in the sun's corona creates solar flares.

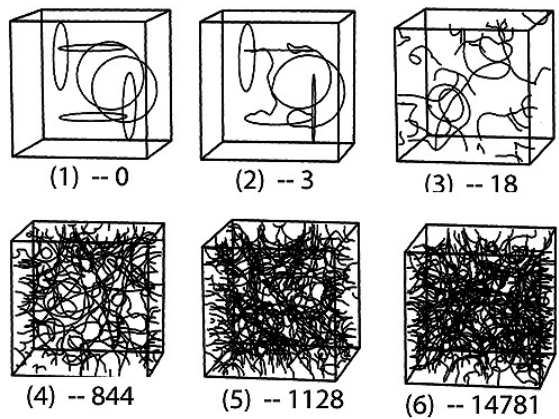

Fig12. Computer simulation of the emergence of quantum turbulence [27]. A heat flux creates a supply of vortex rings, which, through repeated reconnections, become a vortex tangle of fractal dimension. The tangle decays when the heat source is removed. The labels give the frame number and the number of reconnections so far.

We characterize quantum turbulence by the length of vortex lines per unit volume ℓ. Its growth and decay can be described by Vinen's equation

$$\dot{\ell} = A\ell^{3/2} - B\ell^2 \qquad (14)$$

The coefficient A governs the growth of the vortex tangle, and B, its decay. They embody the physical effects of the heat flux that creates large vortex rings, and reconnections that cause their degradation to smaller rings. The balance between these effects determines the size of the steady-state vortex tangle. This equation was postulated by Vinen [28] on a phenomenological basis, derived by Schwarz [27] from vortex dynamics, and experimentally verified in superfluid helium [14].

Our vortices are different from cosmic strings; but apparently, a tangle of cosmic strings can also arise purely geometrically, as singular world surfaces that follow the same equations of motion as the fundamental objects of string theory [29].

## 6. Cosmological equations with quantum turbulence

In constructing a big bang theory, a practical consideration is that we must have spatial uniformity in order to use the Robertson-Walker metric, for otherwise there is little hope of analytical (or even numerical) analysis. To describe quantum turbulence in this context appear to be a hopeless task at first sight, for we have to deal with the phase dynamics of a complex scalar field. The way out is as follows. In the complex scalar field $\phi=Fe^{i\sigma}$, we treat F as spatially uniform, and replace the dynamics of $\sigma$ by that of a uniform distribution of vortex lines as described by Vinen's equation.

The vortex lines are really tubes with finite radius of order of the Planck length. Later, we will estimate the tube radius to be of order $10^{-2}$ Planck length. The space is laced through with vortex tubes, and F is a uniform field in this multiply-connected space. Since there is no scalar field inside the vortex tubes, matter will be created outside of these tubes, and eventually they will form galaxies outside of the vortex tubes.

The independent variables of our system are

$$a(t) = \text{Radius of universe}$$
$$F(t) = \text{Scalar field modulus}$$
$$E_v(t) = \text{Energy of vortex lines}$$
$$E_m(t) = \text{Energy of matter}$$

(15)

The time rate of change a is obtained from Einstein's equation, F from the scalar field equation, $E_v$ from Vinen's equation, and $E_m$ is obtained from the conservation law for the total non-gravitational energy-momentum. The equations of motion are, in units $\hbar=c=4\pi G=1$:

$$\frac{dH}{dt} = \frac{k}{a^2} - 2\left(\frac{dF}{dt}\right)^2 + \frac{a}{3}\frac{\partial V}{\partial a} - \frac{1}{a^3}(E_m + E_v)$$

$$\frac{d^2F}{dt^2} = -3H\frac{dF}{dt} - \frac{1}{2}\frac{\partial V}{\partial F} - \frac{\zeta_0}{a^3}E_v F$$

$$\frac{dE_v}{dt} = s_1 E_v^{3/2} - s_2 E_v^2$$

$$\frac{dE_m}{dt} = -s_1 E_v^{3/2} + s_2 E_v^2 + \zeta_0 \frac{dF^2}{dt}E_v \quad (16)$$

There is a constraint on initial values

$$H = \left(\frac{2}{3}\rho - \frac{k}{a^2}\right)^{1/2}$$

$$\rho \equiv \left(\frac{dF}{dt}\right)^2 + V + \frac{1+\zeta_0}{a^3}E_v + \frac{1}{a^3}E_m \quad (17)$$

which is preserved by the equations of motion. Here, $\zeta_0$ is a constant, and $s_{1,2}$ are parameters that correspond to the coefficients A,B in Vinen's equation (14), and may depend on a(t), and therefore on time. We assume they are of nuclear scale, since they occur in the equation for $E_m$:

$$s_{1,2} \sim \frac{\text{Nuclear energy scale}}{\text{Planck energy scale}} \sim 10^{-18} \quad (18)$$

The smallness of this number decouples the equations into two nearly independent sets, with the first two equations governing the expansion of the universe, and the last two equation describing quantum turbulence and matter creation. The decoupling arises from the fact that the time in the first set is the gravitational time t, whereas the effective time in the second set is the nuclear time

$$\tau = s_1 t \sim 10^{-18} t \quad (19)$$

The matter equation is coupled strongly to Vinen's equation for the vortex system, formally through energy-momentum conservation. These are coarse-grained equations, and the reconnection mechanism is not explicit but implied, as mentioned earlier in the discussion of Vinen's equation.

The cosmological equations have yet to be studied in depth. A scenario of matter creation is proposed in [2] under phenomenological assumptions. The results are shown in Fig.13, where the upper curve shows the growth and decay of quantum turbulence, with inset circles picturing the vortex tangle at various stages. The amount of matter created is the area under the curve. Parameters can be chosen such that the lifetime of quantum turbulence is about $10^{-30}$s, during which time the radius of the universe increases by factor $10^{27}$, and the total amount of matter created is $10^{22}$ sun masses, which will be sufficient for the formation of galaxies.

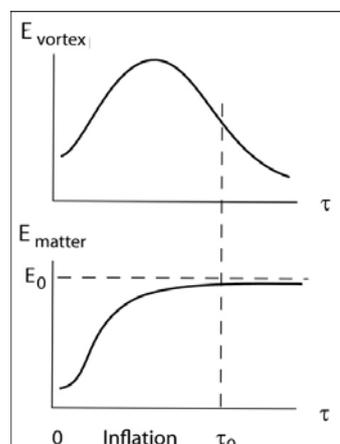

Fig13. The upper curve shows the growth and decay of quantum turbulence, in the form of a vortex tangle, which is pictured in the insets at various stages. Matter is supposed to be created via vortex reconnection, a process essential for the maintenance of the vortex tangle. The amount of matter created is the area under the upper curve.

The lifetime of quantum turbulence corresponds to the inflation era in conventional theory, although there was no special "inflation" in our model, since the universe was always in accelerated expansion. In contrast to usual theories, the scalar field oscillates violently, instead of doing a "slow roll". Matter was created in the vortex tangle instead of "reheating". But what matters is that we can create enough matter in a very short time.

## 7. Post-inflation universe

The time scale in Fig.14 illustrates the very short period of validity of our big-bang model, relative to the very long road ahead to the CMB (cosmic microwave background). At the end of the period of validity, our model passes control to the standard hot big bang model, with one addition: the universe remains a superfluid. To incorporate superfluidity into the standard hot big-bang model would be a huge task. However, we can point out some qualitative manifestations of superfluidity without going into details.

After the demise of the vortex tangle in quantum turbulence, there would be remnant vortex tubes, whose radii will grow with the expanding universe. The radius of a vortex tube at creation was proportional to the radius of the universe a, since that was the only scale available. It will expand with a, maintaining the same ratio to the radius of the universe. Galaxies form outside of the vortex tubes, since there is no scalar field inside. This leaves huge voids in the galactic distribution, with a typical current size of $10^7$ly, or about 1% of the radius of the universe. This would give a tube radius of the order of $10^{-2}$ Planck length at formation.

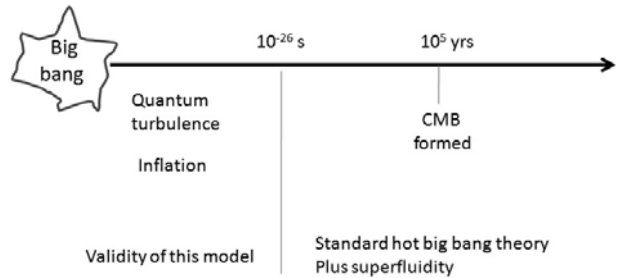

Fig14. Our model ceases to be valid about $10^{-26}$s after the big bang, which marked the end of quantum turbulence, in which all matter in the universe was created. The standard hot big bang model takes over with one addition: the universe remains a superfluid.

The galaxies tend to adhere to the surface of a vortex tube due to hydrodynamic pressure, because the superfluid velocity decreases like $1/r$ from the central axis of the tube. A superposition of vortex tubes in a projected view can reproduce such patterns as the "stick man" [30], as illustrated in Fig.15.

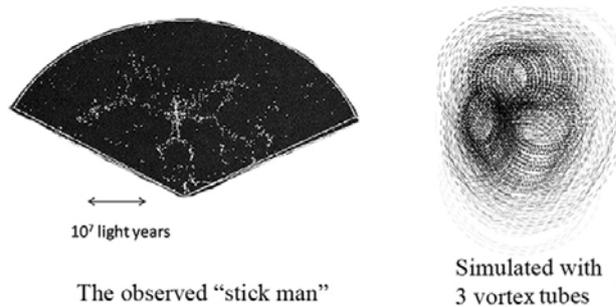

Fig.15. Galaxies form outside of primordial vortex tubes, leaving voids in the galactic distribution. The radii of the vortex tubes were of Planck scale during the big bang, but have since grown to the order of $10^7$ly.

A rapidly rotating body, such a black hole, would drag the superfluid into rotation, with creation of vortices. The black hole would be encaged in vortex lines, as illustrated in the left panel of Fig.16. This could explain [31] the "non-thermal filaments" [32] observed near the center of the Milky Way, as shown in the right panel of Fig.18. The vortex lines can become luminous by trapping electrons inside.

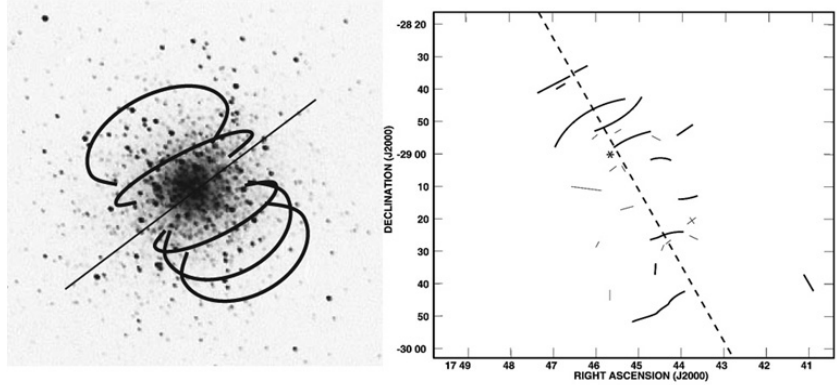

Fig16. A rapidly rotating black hole drags the surrounding superfluid into rotation, by creating vortices, and become encaged in vortex lines, as depicted in the left panel. The "non-thermal filament" observe near the center of the Milky Way, shown in the right panel, could be such vortex lines.

If gigantic remnant vortex tubes should find each other and reconnect, there would appear two jets of intense energy, which could explain some of the observed gamma-ray bursts [33]. They occur at a rate of a few events per galaxy per million years, lasting from ms to minutes, with energy output per second equal to the entire output of the sun in its lifetime (billions of years).

## 8. Dark matter

In the present universe, we can describe superfluidity phenomenologically, using a nonlinear Klein-Gordon equation, with coupling to matter. The high-momentum cutoff can be taken to be constant in time, and we can use any convenient field potential, such as the $\phi^4$ potential. We consider interactions with a "galaxy", a generic reference to any heavenly body, treated as an external source, and focus our attention on the response of the superfluid to the source.

On a similar theme, previous works on Bose-Einstein condensate theory of dark matter may be found in [34-37].

The equation for the complex scalar field in flat space-time is, in units with $\hbar=c=1$,

$$\left(\nabla^2 - \frac{\partial^2}{\partial t^2}\right)\phi - \lambda\left(|\phi|^2 - F_0^2\right)\phi - i\eta J^\mu \partial_\mu \phi = 0 \tag{20}$$

where $J^\mu$ is the mass current of the source. For a rotating galaxy it has the form

$$J^\mu = (\rho, \mathbf{J})$$
$$\mathbf{J} = \rho \mathbf{\Omega} \times \mathbf{r} \tag{21}$$

where $\Omega$ is the angular velocity of the galaxy, and $\rho$ is the mass density of the galaxy taken to be a Gaussian distribution. The current-current coupling between scalar field and external source is dictated by Lorentz invariance. When gravity is taken into account in the Newtonian limit, the equation becomes

$$\nabla^2 \phi - (1 - 2U)\frac{\partial^2 \phi}{\partial t^2} + \frac{\partial U}{\partial t}\frac{\partial \phi}{\partial t} + \nabla U \cdot \nabla \phi - \lambda\left(|\phi|^2 - F_0^2\right)\phi - i\eta J^\mu \partial_\mu \phi = 0 \quad (22)$$

where U is the gravitational potential energy:

$$U(x) = -G \int d^3 x' \frac{\rho_{\text{scalar}}(x') + \rho_{\text{galaxy}}(x')}{|x - x'|} \quad (23)$$

As illustration, we solve these equations numerically in 2D space, with the following results.

In Fig.17, a galaxy is being dragged through the superfluid, which gathers around the galaxy to form a halo (white area) moving with the galaxy like a soliton. This reproduces the dark-matter galactic halo. A transient ripple was created initially. It propagates outward from the center, hitting the computational wall in the last frame.

In Fg.18, two galaxies collide head-on along the vertical direction, and go through each other. Superfluid hydrodynamics governs the contortions of the galactic halos. This may be considered a simulation of the bullet cluster, albeit in 2D.

In Fig.19, two galaxies pass each other at nonzero impact parameter, and apply shear force to the superfluid between them. There develop vortices (black dots) to enable irrotational flow in the superfluid.

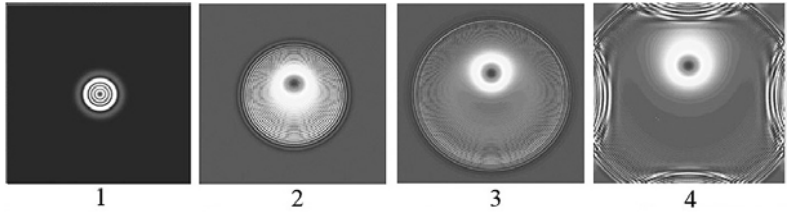

Fig17.   Simulation via the nonlinear Klein-Gordon equation of a galaxy being dragged through the cosmic superfluid. Superfluid is attracted to the galaxy to form the dark-matter halo (white ring), which moves with the galaxy like a soliton. The initial condition creates an outgoing wave in the superfluid, which hits the boundary in the last frame.

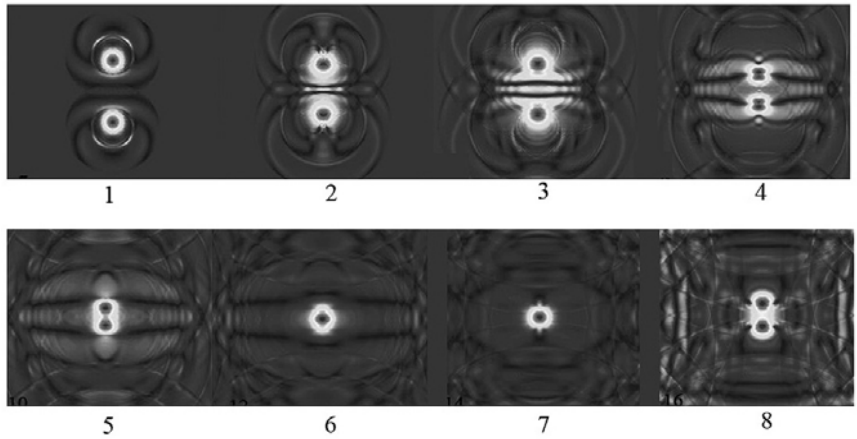

Fig18. Two galaxies collide head-on and go through each other in the cosmic superfluid. The dark matter haloes undergo contortions in accordance with superfluid hydrodynamics.

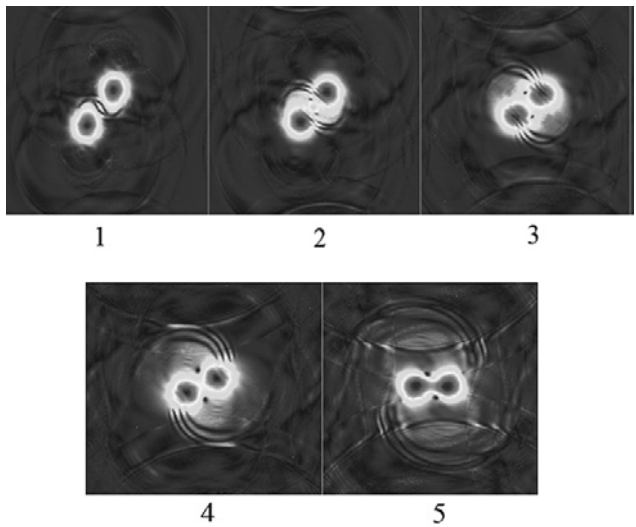

Fig19. Two galaxies collide at nonzero impact parameter. The superfluid is being sheared into rotation, with creation of vortices (the black dots).

A configuration of vortices around a rotating galaxy is depicted in the left panel of Fig.20. From experiments and computations in Bose-Einstein condensates in cold atomic gases [38, 39], we have learned a lot about superfluid vortices. In general, no vortices appear until the angular velocity of the galaxy exceeds a critical value. Then, one vortex would develop at the center of the galaxy. When the angular velocity increases further, new vortices emerge to form a lattice, then the lattice become a ring of increasing radius, and then multiple rings would occur, and this is what is shown in the left panel of Fg.20. The vortices outside the galactic halo (white area) have much large core size, because of increase in correlation length in the superfluid. To verify that the dots in the picture really are vortices, we show a contour plot of the phase of the scalar field in the right panel of Fig.20. The vortices are attached to strings across which the phase of the field jumps by $2\pi$.

Fig.21 shows the number of vortices around a galaxy as function of its angular velocity, for various strengths of coupling η.

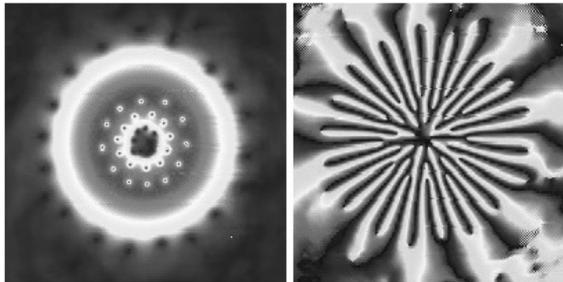

Fig20. The left panel shows the scalar-field modulus around a very rapidly rotating galaxy. There develop 4 rings of vortices, with the one near the center not yet fully formed. The right panel shows the phase of the scalar field. The "spokes" of the wheel are strings across which the phase jumps by $2\pi$.

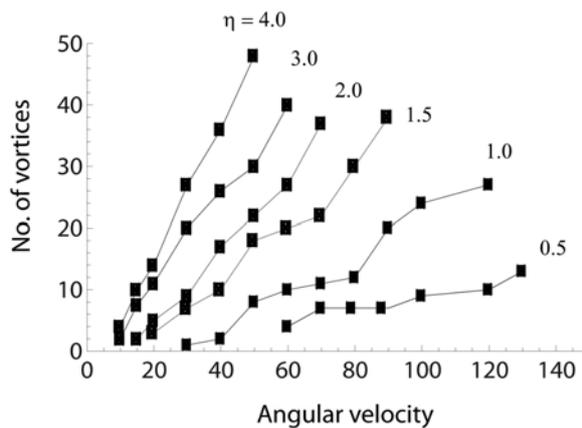

Fig21. Number of vortices around a rotating galaxy, as function of its angular velocity. The parameter η is the coupling strength between galaxy and the underlying superfluid.

In the investigations so far, we work at absolute zero. At a finite temperate, the superfluid will be accompanied by a normal fluid, which consists of phonons and particles that could be excited from the superfluid. For example, the WIMPs being searched for may be coupled to the superfluid, and excitable from it, and this would explain why they are expected to stick around galaxies.

## 9. Conclusion and outlook

The model tries to explain dark energy and dark matter in a unified picture, in terms of a cosmic superfluidity arising from a vacuum complex scalar field. It shows how the quantum scalar field emerges in the big bang, and how all the matter in the universe can be created in quantum turbulence. The results seem encouraging, and give incentive to further developments, possibly along the following lines:

1. incorporation of superfluidity into the standard hot big theory, and analysis of the CMB fluctuations;
2. investigation of black-hole collapse in a cosmic superfluid;
3. clarification of the particle-physics basis for vacuum scalar fields;
4. formulation of the gravitational cutoff to the scalar field through an action principle.

## References

[1] K. Huang, H.B. Low, and R.S. Tung, *Class. Quantum Grav*. **29**,155014 (2012) ; arXiv:1106.5282.
[2] K. Huang, H.B. Low, and R.S. Tung, *Int. J. Mod. Phys*. A **27**, 1250154 (2012); arXiv:1106.5283.
[3] K. Huang, C. Xiong, and X. Zhao, "Scalar-field theory of dark matter", arXiv:1304.1595 (2013).
[4] W. Kolb and M.S. Turner, *The Early Universe* (Addison-Wesley, Redwood City, 1990).
[5] S. Weinberg, *Gravitation and Cosmology*, (Wiley, New York, 1972).
[6] K. Huang, *Quantum Field Theory, from Operators to Path Integrals*, 2nd ed (Wiley-VCH, Weinheim, Germany, 2010).
[7] K. Huang, *Quarks, Leptons, and Gauge Fields*, 2nd ed (World Scienific, Singapore, 1992).
[8] A. G. Riess et al., *Astrophys. J*. **659**, 98 (2007).
[9]  V. Rubin and W. K. Ford Jr., *Astrophys. J.* **159**, 379 (1970).
[10] D. Clowe et al, *Astrophys. J*. **648**, L109 (2006); arXiv:0608407.
[11] M.Tegmark et. al., *Phys. Rev*. D **69**, 103501 (2004); arXiv:0310723.
[12] PICASSO Collaboration*, Phys. Letters* B **682**, 185 (2009); arXiv:0907.0307.
[13] V.L. Ginsburg and L.D. Landau, *Zh. Eksperim. i. Teor. Fiz*. **20**, 1064 (1950).
[14] S.K. Nemirovskii and W. Fizdon, *Rev. Mod. Phys*. **67**, 37 (1995).

[15] CMS collaboration, *Phys. Letters* B **716**, 30 (2012).

[16] J.Bardeen, L. Cooper, and J.R. Schrieffer, *Phys. Rev.* **106**,162 (1957).

[17] F. Dafolvo, S. Giorgini, L.D, Pitaevskii, and S. Stringari*, Rev. Mod. Phys*. **71**, 463 (1999).

[18] F.J. Dyson, *Phys. Rev*. **75**, 486 (1949).

[19] K.G. Wilson, *Rev. Mod. Phys*. **55**, 583 (1983).

[20] Chap.16., Ref. [6].

[21] K. Halpern and K. Huang, *Phys. Rev*. **53**, 3252 (1996).

[22] V. Periwal, *Mod. Phys. Lett.* A **11**, 2915 (1996).

[23] Chap.17, Ref.[6].

[24] M.E. Peskin and D.V. Schroeder, *An Introduction to Quantum Field Theory (Westview Press, Boulder, 1995).*

[25] J. Barlow, A.R. Liddle, and C. Palud, *Phys. Rev*. D**74**, 127305 (2006).

[26] M.S. Paoletti, M.E. Fisher, K.R. Sreenivasan, and D.P. Lathrop, *Phys. Rev. Lett.* **101**, 154501 (2008).

[27] K.W. Schwarz, *Phys. Rev.* **B** 38, 2398 (1988).

[28] W.F. Vinen, *Proc. Roy. Soc. London*, A **114** (1957); **240**, 128 (1957); **243**, 400 (1957).

[29] H. Kleinert, *EJTP*, **8**, 27 (2011); arXiv:1107.2610.

[30] V. deLapparent, M.J. Geller, and J.P. Huchra, *Astrophys. J*. **302**, L1 (1986).

[31] D.P. Lathrop (private communication).

[32] T.N. LaRosa et. al*., Astrophys. J*. **607**, 302 (2004).[31]

[33] *Gamma-Ray Bursts*, K. Chryssa, E.W. Stanford, A.M.J. Ralph, eds (Cambridge University Press, Cambridge, England, 2012).

[34] J.-W. Lee, S. Lim, and D. Choi, arXiv:0805.3827 (2008).

[35] J.-W. Lee, J. *Korean Phys. Soc*. **54**, 2622 (2009); arXiv:0801.1442.

[36] B. Kain and H.Y. Ling, *Phys. Rev*. D**82**, 064042 (2010); D85, 023527 (2012).

[37] A. Suarez, V.H. Robles, and T. Matos, arXiv:1302.0903 (2013).

[38] M. Tsubota, K. Kasamatsu, and M. Ueda*, J. Low Temp. Phys*. **126**, 461 (2002).

[39] A.L. Fetter, *Rev. Mod. Phys*. **81**, 647 (2009).